\documentclass[12pt]{iopart}
\usepackage{iopams}
\usepackage{graphicx}
\usepackage{epsfig}
\usepackage{color}

\begin{document}

\title
{High threshold distributed quantum computing with three-qubit nodes}

\author{Ying Li}
\address{Centre for Quantum Technologies, National University of Singapore, 3 Science Drive 2, Singapore 117543}
\author{Simon C. Benjamin}
\address{Department of Materials, University of Oxford, Parks Road, Oxford OX1 3PH, UK}
\address{Centre for Quantum Technologies, National University of Singapore, 3 Science Drive 2, Singapore 117543}

\begin{abstract}
In the distributed quantum computing paradigm, well-controlled few-qubit  `nodes' are networked together by connections which are relatively noisy and failure prone. A practical scheme must offer high tolerance to errors while requiring only simple (i.e. few-qubit) nodes. Here we show that relatively modest, three-qubit nodes can support advanced purification techniques and so offer robust scalability: the infidelity in the entanglement channel may be permitted to approach 10\% if the infidelity in local operations is of order 0.1\%. Our tolerance of network noise is therefore a order of magnitude beyond prior schemes, and our architecture remains robust even in the presence of considerable decoherence rates (memory errors). We compare the performance with that of schemes involving nodes of lower and higher complexity.
Ion traps, and NV- centres in diamond, are two highly relevant emerging technologies.

\end{abstract}

\maketitle

\section{Introduction}

Over the last two decades a remarkable range of physical systems have been shown to support the key elements required for quantum computation: robust qubits, controlled single qubit gates and measurements, and entangling operations. However these successes have involved relatively few qubits in any given device. It remains to be seen whether the high fidelity operations and low decoherence rates that have been achieved in small systems can be replicated in vastly more complex arrays of thousands, or millions of qubits. 
There is however an attractive alternative approach to creating monolithic arrays, namely {\em distributed quantum computing} (DQC): one can form a network of small devices or `nodes', such that nodes are well separated and therefore each can have dedicated control and measurement systems similar to those that have already been achieved. The challenge is then to entangle the `remote' systems (remote here meaning well separated but still part of the same physical device).  Experiments  with atoms \cite{D.L.Moehring2007} and atomic ensembles \cite{C.W.Chou2005} have shown that two remote quantum entities can undergo an entanglement operation (EO) when light emitted (or potentially emitted) from the structures is suitably measured, and solid state demonstrations may soon follow~\cite{hanson2012}. Generally the network EO may be of low fidelity and may fail outright much of the time, and this must be accounted for in practical DQC schemes. 

In a typical DQC scheme, one assumes that (at least) one qubit in each node is directly operated upon by the network EO -- this is how `raw' entanglement is generated between nodes. Typically it is presumed that the basic EO involves only two nodes at a time (although schemes for entangling multiple nodes in one step have been advanced \cite{Benjamin2005NJP, Fujii2011}). However one assumes that multiple such pairings can be attempted simultaneously, i.e. there is parallelism in the overall architecture. When a network EO fails, the corresponding qubit in each of the two nodes will be corrupted. Here we will not need to specify further  physical details of the EO process; it may involve `path erasure' techniques~\cite{SeanD.Barrett2005, YuanLiangLim2005, D.L.Moehring2007R} or alternatively continuous variable mediators as in the `qubus'~\cite{bill2011} approach, or even recently advanced ideas involving combining many low-level couplings~\cite{Borneman2012}. We require only that there is {\em some channel} by which entanglement, possibly of low quality, can be generated between nodes.

The robustness to errors in DQC architectures is highly dependent on the size of the nodes: nodes with a higher number of qubits can support more complex error suppression techniques. On the other hand, it is desirable to minimise the number of qubits per node in order to reduce the experimental challenge and support the widest range of physical implementations. In the following we will write DQC-$N$ to refer to schemes for distributed QIP involving nodes with $N$ qubits. Note that the acronym DQC1 has been used in the literature relating to the `power of one pure qubit' but this is not the meaning intended here.

\section{The states of the art}

DQC-1, i.e. networked QIP with only one qubit per node, is in principle possible. In that case one finds that the rate of unknown errors in the network EO must be low if the overall device is to support scalable QIP, however one can at least tolerate a high rate of known (`heralded') failures: Refs. \cite{YingLi2010} and \cite{Fujii2010} describe tolerating $~90\%$ failure rates using a resource overhead of the order of a factor of a thousand. This overhead is required because each EO failure damages the effort to create large scale entanglement, necessitating massive redundancy. 

In DQC-2, the two qubits can be purposed for different tasks: one qubit in each node to be used in the network EO, and one qubit to build up multi-qubit entanglement with other nodes. By separating these tasks, we at least gain the advantage that heralded failures in the EO can simply be repeated without the cost of losing previously achieved entanglement. The term ``broker'' has been used for the qubit that is involved in the raw network EO, while ``client'' refers to the qubit that receives the successful entanglement from the broker \cite{SimonCBenjamin2006}. While the broker/client model allows one to absorb heralded failures without the massive overhead of DQC-1, the node size is still too small to be useful in removing general noise (i.e. unknown errors) on the internode entanglement. Thus, the fidelity of successful EOs must still be high in order to support full scale QIP (strictly speaking, we cannot handle {\em full rank} noise, however if for example the noise is primarily phase noise, then strategies are available -- we return to this point later). 

For a practical machine is highly desirable to be able to {\em purify} internode entanglement: in effect to use multiple `raw' entanglement operations, each prone to noise, to eventually produce a single higher fidelity internode operation.  Thus one can support physical systems where the native EO fidelity is far below the levels required for conventional fault tolerant QIP. In 2003 Briegel and D\"{u}r showed that entanglement purification can be efficiently performed with a limited number of ancilla qubits; they considered DQC-3, DQC-4 and DQC-5~\cite{Dur2003}. This seminal work established `the bar' for distributed QIP with different node sizes, and until the present paper it remained the best approach for DQC-3. However for larger nodes there has been subsequent progress: for example, DQC-5 was examined in considerable detail in 2007 by Jiang {\em et al}~\cite{LiangJiang2007}, where the authors demonstrated that errors in the network EO, {\em and} in both measurement and preparation can reach $5\%$. The caveat here is that error rates on local gates are required to be very low, $0.01\%$ or less. A recent a preprint from Fujii {\em et al} has again advanced the state of the art for DQC performance: they detail a DQC-4 scheme which supports fully fault tolerant QIP given a local gate error rate of $0.1\%$ and a network EO fidelity as low as $70\%$ \cite{Fujii2012}. The authors neglect memory errors (i.e. decoherence from sources other than gate operations) but note that this will have a limited impact. 

Here we introduce and analyse an advanced DQC-3 scheme capable of fault tolerant QIP despite high levels of full rank noise on the network EO, along with significant errors rates in the local (intra-node) operations and memory errors. We find that our approach outperforms the previously established DQC-3 protocol, tolerating network noise that is an order of magnitude more severe. Our approach achieves this by `cramming in' two levels of purification within the limited space of three qubits. This comes at the cost that arbitrary inter-node two-qubit operations are not supported. However, fortunately such gates are not required and in fact one particular kind of operation, the parity projection, can suffice to generate the entangled resource required for topologically protected QIP. This is the key insight for the novel DQC-3 protocol we describe below. 

\section{Overview of full noise purification with DQC-3}

Here we establish that three qubits per node can suffice to enable distributed quantum computing with very noisy network EOs. The possibility of  purifying highly imperfect network operations using only DQC-3 was mentioned by Campbell in 2007, when he developed the concept of purifying a parity projection {\em operation} rather than purifying a quantum state~\cite{Campbell2007}. Campbell showed that, given a source of entangled pairs whose infidelity is largely phase noise (i.e. bit noise is low), one can directly perform a high fidelity parity projection (PP) on two client qubits in an unknown state. Here we incorporate Campbell's approach: We pair up the single intermediate qubit in each of two nodes, purify bit errors from the brokers onto this pair, and since this pair now exhibits principally phase noise it can use used to direct apply a PP to the clients. 

In effect Campbell's technique repeatedly performs a PP until one gains sufficient confidence that it has occurred with outcome either `odd' or `even' -- this exploits the fact that once a PP has occurred, successive PPs have no new effect on the qubits (in contrast to, say, the controlled-phase gate which would negate the effect of a prior application). Here we show how to use PPs to build a particular large scale entangled state between our nodes: namely Raussendorf's topologically protected cluster state (TPCS) \cite{R.Raussendorf2006, RobertRaussendorf2007, R.Raussendorf2007}. This is the same state that is the target of Refs. \cite{YingLi2010, Fujii2010, Fujii2012}. It is a convenient target since one need only show that the state can be built (and measured) with a certain fidelity in order to assert that universal QIP is then possible. Here, we need to find a route to generating the TPCS with the fewest possible parity projections in order to minimise the accumulation of errors. The route we employ is discussed below; in obtaining our threshold we also benefit from work by Barrett and Stace \cite{SeanD.Barrett2010} which shows that the TPSC need not be perfect in order to retain its fault tolerance (up to $25\%$ of the structure can be deleted provided of course that one knows which parts are absent).

\section{Purifying the parity projection operation}\label{distilSection}

We wish to perform a high fidelity parity projection operation on the client qubits in each of two separate nodes of our DQC-3 machine. Each node contains a qubit which is directly associated with the (noisy) entanglement channel, i.e. the broker qubit -- for example, an electron spin with an associated conditional optical transition which therefore permits non-local entanglement via `path erasure' type schemes \cite{SeanD.Barrett2005, YuanLiangLim2005, D.L.Moehring2007R}. Additionally each node contains one intermediate qubit. We assume that the primal network entanglement operation (EO) on the two  broker qubits should ideally result in one of the four Bell states
\begin{equation}
\left\vert {\rm bell} \right\rangle=\frac{1}{\sqrt{2}}[\left\vert 0,\mu \right\rangle +(-1)^{\nu }\left\vert 1,\mu \oplus 1\right\rangle ],
\end{equation}
where $\mu ,\nu =0,1$, but that in practice this state may be affected by errors. Without loss of generality we may write the erroneous superoperation as
\begin{equation}
E_{\mathrm{ent.}}=A[1]+B[Z]+C[X]+D[Y]. \label{Eent}
\end{equation}
where $A+B+C+D=1$, $A$ is the fidelity of the entanglement and Pauli errors are acting on arbitrary one of the two qubits. Here and throughout we use the form $E = \sum_{j}  [U_{j}]$
to denote the superoperator
$E( \rho ) = \sum_{j} U_{j} \rho U_{j}^{\dagger}$.

\begin{figure}[h]
\centering
\includegraphics[width=13cm]{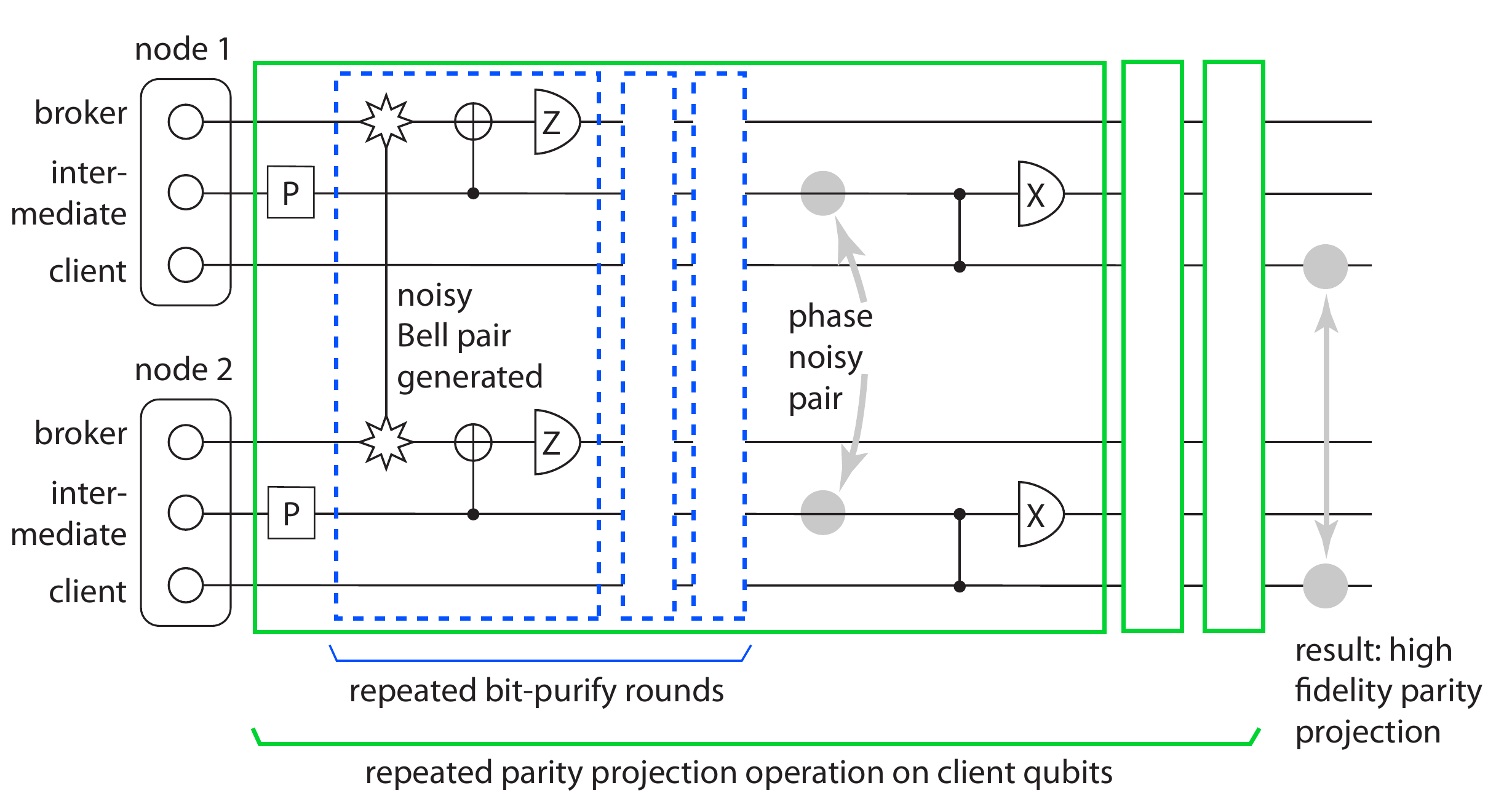}
\caption{
The full protocol for implementing a high fidelity parity projection on the clients. Note that only the brokers are involved in an operation spanning the two nodes: this is the network EO which generates a noisy Bell pair. The box marked P denotes preparation of each intermediate qubit into a state such as $|+\rangle$. In a typical physical implementation the broker qubit is an optically active system whereas the other two qubits in each node need not be (they might be realised as nuclear spins, for example). Preparation and measurement of the intermediate qubit could  in practice be achieved by transferring the state to/from the broker for optical readout. The client qubits  of course never undergo preparation or measurement in this process: the parity projection applies to an arbitrary incoming client state.
}
\label{protocol}
\end{figure}

We employ a bit-error entanglement pumping scheme according to which we generate a succession of `raw' Bell states on the broker pair, and via local gates and measurements we progressively generate a state on the intermediate qubit pair with an improved noise profile:  flip errors are heavily suppressed, although phase errors will have increased~\cite{LiangJiang2007}.

We now use this intermediate pair to perform a parity projection on the client qubits (see Fig.~\ref{protocol}, outer level of nesting). For a moment, let us neglect the significance of residual flip errors on the intermediate qubit pair, and local gate errors. Consider the consequence of the substantial phase error probability $p_P$: in the event of such an error one finds that a PP is nevertheless performed successfully on the clients, but we have an incorrect record of whether the outcome was `odd' or `even'. However the intermediate qubits are now available again, and we are free to repeat the previous pumping procedure and generate a new bit-flip-corrected pair, and so perform a second parity projection on the clients. Indeed we can repeat this as many times as we wish. Campbell's observation is that we can then deduce the likely true parity from looking at the relative frequency with which `odd' and `even' are reported.

After $h=m+2l$ parity projections on a pair of client qubits, where $m+l$ measurement outcomes indicate the same parity while $l$ indicate the other parity, the probability of a wrong parity is reduced to
\begin{equation}
P_{\mathrm{P}}=\frac{\alpha ^{-m}}{\alpha ^{m}+\alpha ^{-m}},
\end{equation}
where $\alpha =\sqrt{p_{\mathrm{P}}^{-1}-1}$.
The next effective parity projection increases the difference $m$ with a probability
\begin{equation}
h_{m}=\frac{\alpha ^{m}(1-p_{\mathrm{P}})+\alpha ^{-m}p_{\mathrm{P}}}{\alpha ^{m}+\alpha ^{-m}}.
\end{equation}

If there were no types of error other than phase errors on the source, then one could continue this process to produce an arbitrarily high fidelity parity projection (i.e. continue until difference $m$ is very large). In practice, there are residual flips errors on the source, errors on local operations within each node, and memory errors. The effect of these errors increases with the number of effective parity projections $h$.
Therefore, it is useful to truncate $m$ and $h$.
We introduce the rule that if the difference $m$ is at least some target $M$ after at most as $H$ parity projections on the same pair of client qubits, then the parity projection is deemed successful. Otherwise the protocol is deemed a failure. This means that we will have (known) imperfections in the TPCS that we are trying to generate; but the work of Barrett and Stace~\cite{SeanD.Barrett2010} allows us to translate this damage rate into an lowered threshold at which the TPCS achieves fault tolerance. This tradeoff between PP fidelity and the threshold of the TPCS gives us an opportunity numerical optimisation which we exploit in obtaining our results.

\section{Overview of topological fault-tolerant quantum computing}

\begin{figure}[tbp]
\centering
\includegraphics[width=9cm]{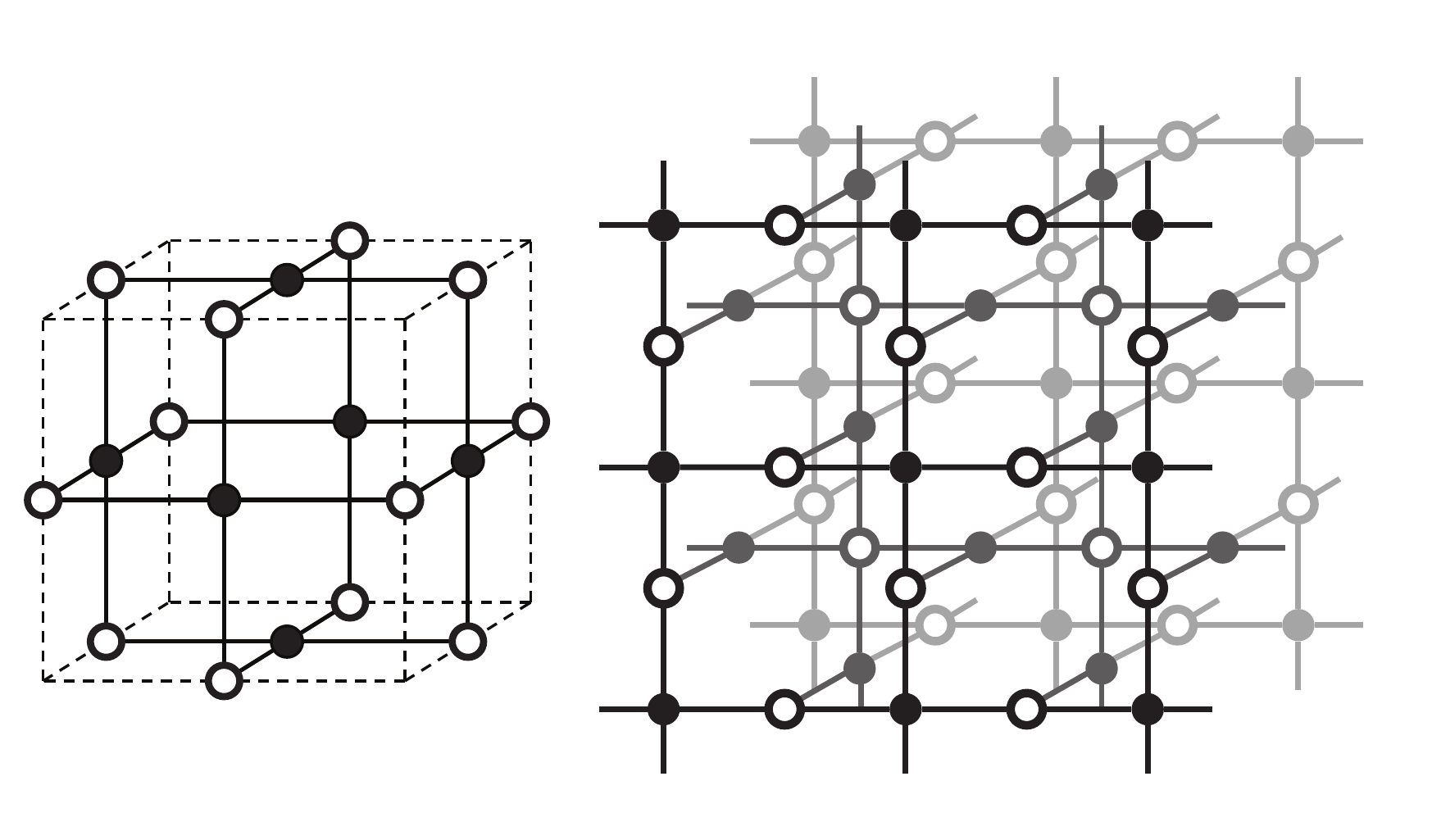}
\caption{
Left: The elementary cell of the topologically protected cluster state. 
Right: This cell repeats to generate a large 3D structure (which can however be flattened to quasi-2D or true 2D). 
Here open/filled circles are a guide to the eye showing that one may understand the structure as two interlaced lattices: these are called the primary lattice and the dual lattice.
}
\label{TFTQC}
\end{figure}

Here we describe the relevant features of the topologically protected cluster state (TPCS) that we aim to generate. Raussendorf and his collaborators introduced the TPCS in Refs. \cite{R.Raussendorf2006, RobertRaussendorf2007, R.Raussendorf2007}. The elementary cell of the TPCS is shown in Fig. \ref{TFTQC}, where the diagram is to be understood to depict a graph state. Any such diagram be easily understood through the so-called constructive definition: suppose that each vertex represents a qubit initially prepared in the $\left | + \right \rangle$ state, then each connecting line or `edge' represents a control-phase gate between the corresponding two qubits. The Figure represents the physical state that would result from following this prescription, therefore obviously one way to generate the TPCS is to do just that. However here we do not have the capability to perform control-phase gates but rather only the parity projection. Therefore we need to determine how to construct the same final state using the PP as our primitive. This is discussed in the next section.

The remarkable property of the TPCS is that, having created it, one can use it as a resource that enables universal quantum computation -- no further entangling operations are required, in fact one only needs to perform single qubit measurements of each vertex, effectively consuming the TPCS as the computation proceeds. This property depends on the noise on the TPCS being below a certain threshold, as described below. Provided that that we can show a given DQC scheme can indeed generate the TPCS with acceptable levels of noise, ``we are done'' in the sense that any quantum algorithm can now be performed by measuring individual client qubits in the fashion specified by Raussendorf {\em et al.}

\section{Building the TPCS within the constraints of DQC-3}

\begin{figure}[h]
\centering
\includegraphics[width=16cm]{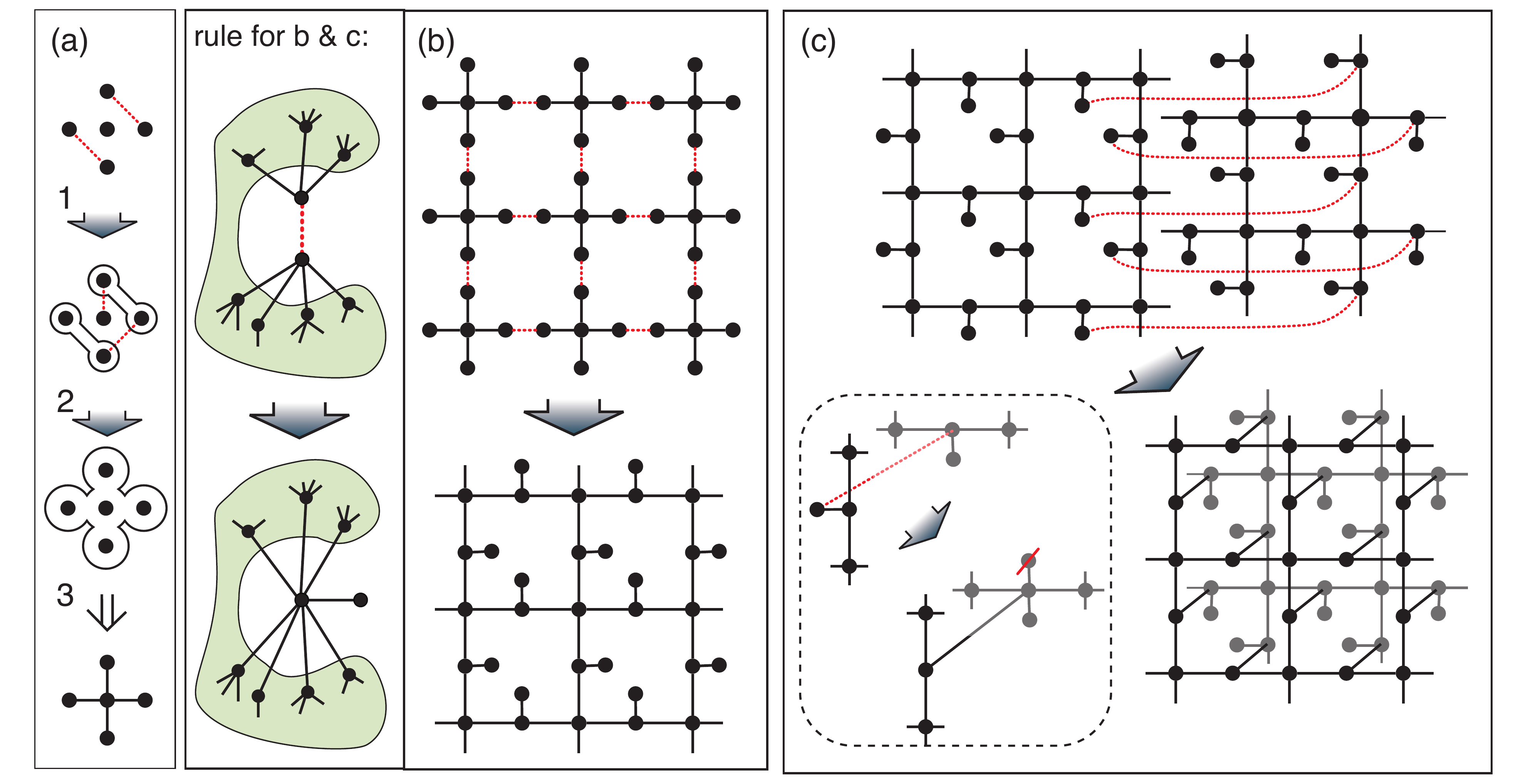}
\caption{
The strategy we adopt for generating the full TPCS. The graph state notation is employed. Red dashed lines indicate pairs of qubits that will be subject to a PP in the next step. The wide, shaded arrow indicates that PPs are applied, and possibly local operations (often these can be omitted or subsumed over several steps).  In (a) we show how to generate a `cross' graph: the first two steps each involve two parallel PPs, the third step is simply a set of local rotations. In the next panel we show a general graphical rule for applying a PP plus local gate(s) on two qubits that are already part of a graph state: after the PP we rewrite the graph with an arbitrary one of the two qubits inheriting all their prior connections, and the other as a `dangling bond'. In (b) we connect a grid of cross graphs into a lattice (unterminated lines indicate the lattice can continue to arbitrary size with the same pattern). In (c) we take two such lattices and connect them so that they become two planes of the TPCS. The inset shows that each PP is followed by removing one qubit in the $z$ basis. Note that the final object has dangling bonds in the more distant plane, and these are exactly as required to repeat the process and connect another lattice sheet (and in fact multiple sheets may be fused in one time step).
}
\label{growingTheState}
\end{figure}
In Section~\ref{distilSection} we described how DQC-3 enables high fidelity parity projection (PP) between nodes of the computer. We must now harness this entangling operation for generating the target TPCS. It is essential to use as few operations as possible since each PP will introduce some risk of errors to the structure. The procedure we employ is depicted in Figure~\ref{growingTheState}. 

The process is quite efficient: We require only five time steps in order to generate a two dimensional lattice as shown in panel (b): Specifically one step to prepare client qubits, three further steps in panel (a), and then a single PP step in panel (b). We then need only two further steps (one PP, one measurement) to combine any number of such lattices into the full TPCS. This is comparable to the number of steps that would be required to generate a TPCS by directly following the constructive definition, i.e. using controlled-phase gates (which require at least DQC-4 if purification is to be employed). Consequently the thresholds which we presently determine for DQC-3 are not very much lower than those which DQC-4 attains.

\section{Determining the thresholds for fully scalable QIP}

In order to establish the thresholds for fault tolerant QIP using DQC-3, we must translate from the physical qubit errors that will occur during the PP operations and local rotations/measurements, to an effective error rate afflicting the resulting TPCS. If this latter error rate is below a certain threshold, then fault tolerant QIP will be possible.

According to previous studies, the error rates that the TPCS can tolerate depends on the classical algorithm used to analyse the results of measurements (i.e. detecting defects and matching them). Two algorithms that have been studied are the {\em minimum-weight perfect matching algorithm} \cite{R.Raussendorf2006} and  the {\em random plaquette $Z_{2}$-gauge model} \cite{RobertRaussendorf2007}. In this paper, we focus on the {\em minimum-weight perfect matching algorithm}.

The error-rate threshold is usually obtained by modelling imperfect local operations with depolarizing errors.
An imperfect operation with depolarized errors is described as a combination of a perfect operation and an erroneous superoperation,
\begin{equation}
E_{1}=(1-p)[1]+\frac{p}{3}([X]+[Y]+[Z])  \label{E1}
\end{equation}
for single-qubits operations (e.g. single-qubit gates and measurements) and
\begin{equation}
E_{2}=(1-p)[1]+\frac{p}{15}([I_{1}X_{2}]+\cdots +[Z_{1}I_{2}]+\cdots +[X_{1}Y_{2}]+\cdots ) \label{E2}
\end{equation}
for two-qubit operations (e.g. controlled-NOT gate, controlled-phase gates), where $p$ is the error rate.
Importantly, in the TPCS any error is equivalent to a phase error (or correlated phase errors) by noticing that the cluster state is stabilized by $K_{a}=X_{a}\prod_{b\in N_{a}}Z_{b}$, where $N(a)$ is the set of four qubits neighboring the qubit-$a$.
Error correction of qubits on faces (filled circles in Fig. \ref{TFTQC}) and qubits on edges (open circles in Fig. \ref{TFTQC}) of the cubic lattice are performed independently.
Therefore, correlations between these two sets of qubits can be neglected.
After neglecting these correlations, phase errors are independent or occurring in the correlated form $[Z_{1}Z_{2}]$ on two qubits connected with the same qubit.
We assume the {\em minimum-weight perfect matching algorithm} is employed with performance as described in~\cite{R.Raussendorf2006}.
The low-level threshold corresponds to independent phase errors ($[Z_1]$, $[Z_2]$) on the TPCS with the probability $2.34\%$ and correlated phase errors ($[Z_1Z_2]$) with the probability $0.59\%$. Because the threshold of totally independent phase errors is $2.93\%=2.34\%+0.59\%$ \cite{ChenyangWang2003}, we can conclude that by using the {\em minimum-weight perfect matching algorithm}, the presence of correlations in the noise does not affect the threshold significantly.

Fault tolerant QIP on the TPCS can tolerate not only unknown errors, but also outright qubit loss.
Numerical evidence suggests that the error threshold decreases approximately linearly with the probability of qubit loss and qubit loss less than $24.9\%$ is tolerable \cite{SeanD.Barrett2010}.
In this paper, we neglect all correlations in phase errors and estimate error thresholds with
\begin{equation}
\epsilon _{t}=2.93\%-\frac{2.93\%}{24.9\%}p_{\mathrm{loss}},  \label{ET}
\end{equation}
where $\epsilon _{t}$ is the probability threshold of phase errors on the TPCS, $p_{\mathrm{loss}}$ is the loss rate.

\section{Results}

\begin{figure}[bp]
\centering
\includegraphics[width=12cm]{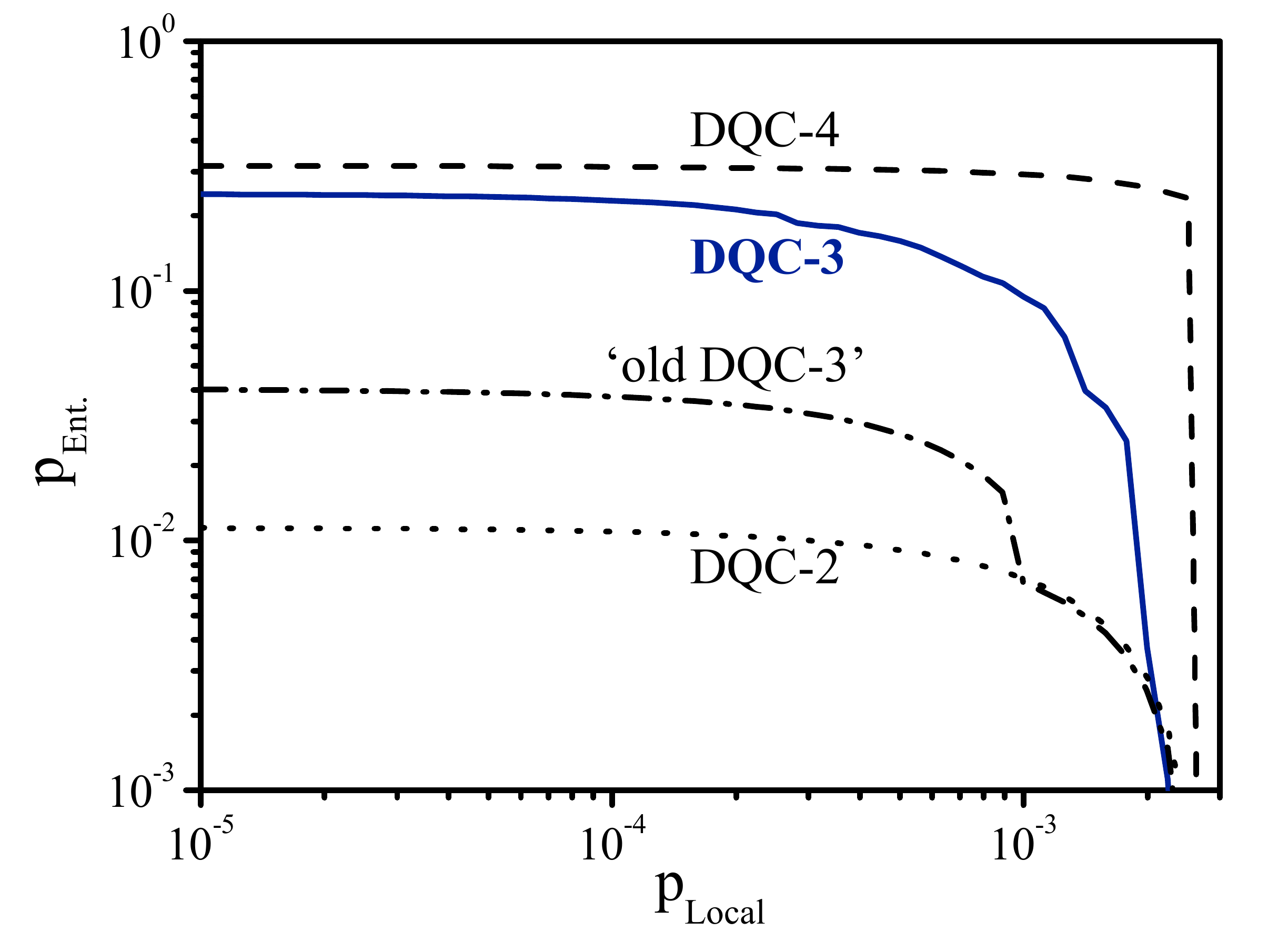}
\caption{State of the art thresholds for fully fault tolerant distributed quantum computing. The probability of an error on the `raw' network entangling operation is $p_{Ent.}$ and the equivalent probability for local operations within a node (rotations, two-qubit gates or measurements) is $p_{Local}$. The area to the lower left of any line represents the region in which fault tolerant QIP is possible. Here memory errors are neglected (c.f. Fig.~\ref{thresholdsM}). The DQC-3 performance (blue) is derived from the analysis in this paper. For comparison we plot also the DQC-4 performance which we reproduce from a graph in Ref.~\cite{Fujii2012}. We note that very high rates of infidelity can be tolerated by DQC-3 although the additional qubit in DQC-4 does allow still greater tolerance (at local error rates of $10^{-3}$ tolerance of network infidelity approaches $10\%$ for DQC-3 and $30\%$ for the DQC-4). We also plot the ``previous best'' protocol for DQC-3, as described in the text.}
\label{keyResultsFig}
\end{figure}

\begin{figure}[tbp]
\centering
\includegraphics[width=12cm]{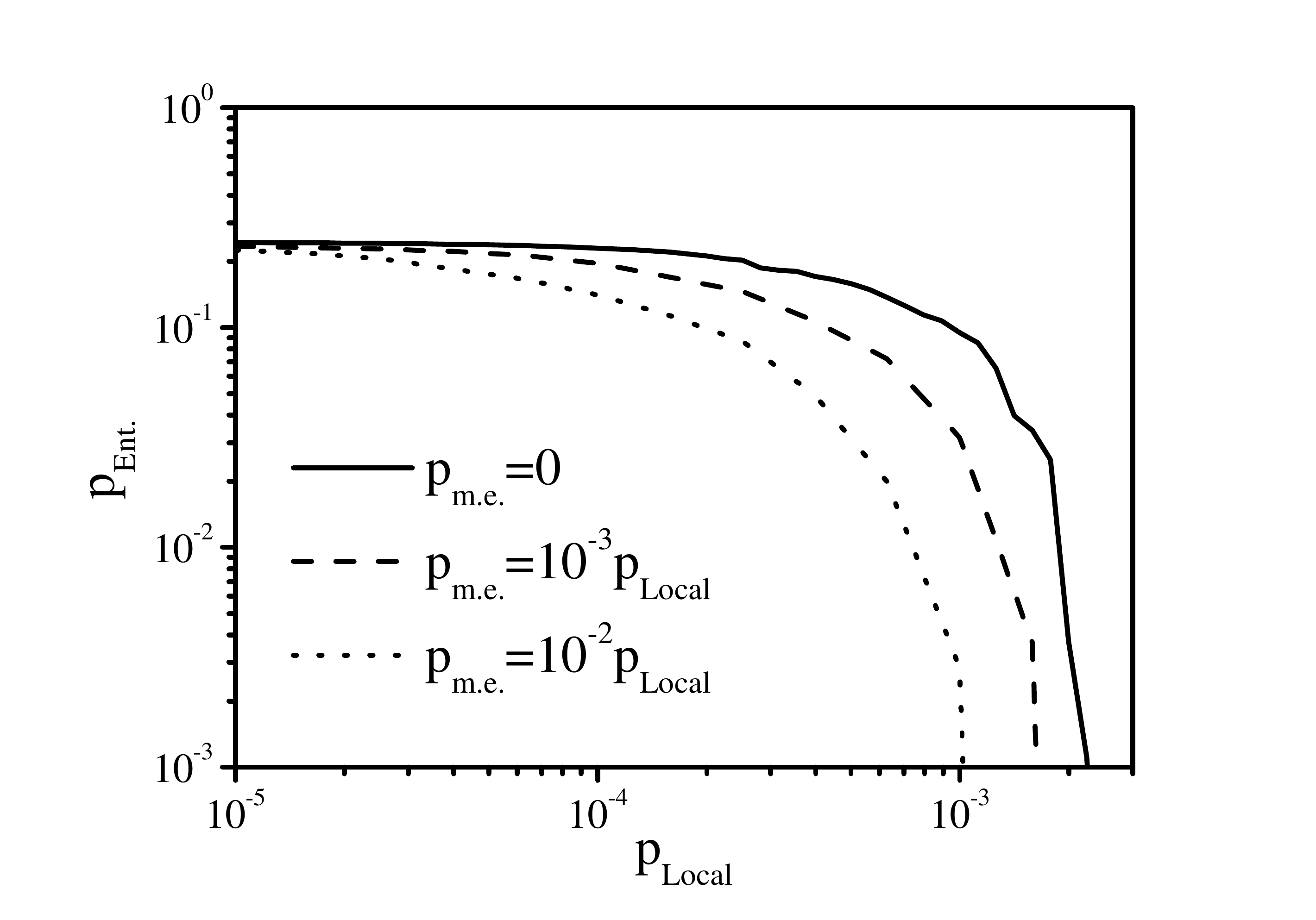}
\caption{This plot indicated the impact of memory errors. Here $p_{mem}$ is the probability that an error occurs on an inactive qubit during one typical time step of the basic protocol. Rates are as given in the figure; in all cases we assume a $90\%$ rate of heralded EO failure (see text). 
}
\label{thresholdsM}
\end{figure}

Using the techniques described in the preceding sections, we obtain conditions for fault tolerant QIP in terms of the fundamental characteristics of the DQC-3 machine: the fidelity on the network EO, the fidelity of local operations, and the memory error rate (i.e. the rate at which errors occur even when no operations are applied). There are also parameters internal to the protocol, such as the choice of how many rounds to allow for purifying the parity projection before abandoning it -- these parameters are selected by numerical optimisation. 

Our results are shown in Figures~\ref{keyResultsFig} and \ref{thresholdsM}. We make the simplification that all local errors, i.e. rotation, local entanglement and measurements, occur with the same probability. In Fig.~\ref{keyResultsFig} we plot the case where memory errors can be neglected, and for contrast in Fig.~\ref{thresholdsM} we plot the case of significant memory errors. Figure~\ref{keyResultsFig} includes a line which shows the performance of the original DQC-3 scheme described in Ref.~\cite{Dur2003}. Until the present paper, this was the most efficient protocol for three-qubit nodes. The curve  labelled `old DQC-3' is derived by taking the purification protocol of Ref.~\cite{Dur2003} and using it to implement internode phase gates so as to directly create the TPCS. Note that our scheme outperforms the original DQC-3 protocol quite dramatically: for low rates of local error, we tolerate about six times more internode noise, and for local gate errors in the region of $10^{-3}$ this factor reaches ten.

As an aside we note that the performance of our DQC-3 scheme actually closely matches that of the original DQC-4 approach in Ref.~\cite{Dur2003}. In fact this makes intuitive sense: both schemes use a two tier entanglement purification (ours `crams it all in' to DQC-3 by purifying the parity projection {\em operation}, rather purifying a Bell state that then empowers a subsequent operation).

Note that the memory error rates chosen for Fig.~\ref{thresholdsM} are more severe than in many real systems: for example it may take only tens of microseconds to rotate a nuclear spin, whereas the spin decoherence time can be of order minutes~\cite{johnMemory}.
We also apply a very high outright (heralded) failure rate to the network EO of $90\%$. This means that each time the protocol requires a new `raw' entangled pair to be generated, on average there will be 10 attempts required before the protocol can proceed. Even under these highly adverse conditions, the DQC-3 scheme still tolerates network EO infidelity of order $10\%$ for local error rates well above $10^{-4}$.

\section{Conclusions}

Here we describe an approach to distributed quantum computing using a machine with three qubits in each of its nodes. We show that even such a limited system can offer robustness through the process of purifying the infidelity on the network (i.e. internode) entangling operation. While our DQC-3 protocol cannot quite match the performance of the recently described DQC-4 procedure \cite{Fujii2012}, the significant saving in node complexity may be attractive to experimentalists. For example, an NV- centre with an optically active electron state and two coupled nuclear spins (say, one nitrogen and one carbon-13 nucleus) can suffice to realise the protocol described here.

We thank Earl Campbell for helpful discussions. This work was supported by the National Research Foundation and Ministry of Education, Singapore.

\end{document}